 \documentstyle[prl,aps,epsf]{revtex} 

\begin{document}

 \twocolumn[\hsize\textwidth\columnwidth\hsize\csname@twocolumnfalse%
 \endcsname 

\draft


\title{ \vskip -0.5cm
         \hfill\hfil{\rm\normalsize Printed on \today}\\
         Photogalvanic Effects in Heteropolar Nanotubes}

\author{Petr Kr\'al$^{1}$, E. J. Mele$^{2}$
and David Tom\'anek$^{3}$}

\address{$^{1}$ Department of Chemical Physics, Weizmann Institute of Science,
         76100 Rehovot, Israel}

\address{$^{2}$ Department of Physics, Laboratory for Research on the 
Structure of Matter,\\
University of Pennsylvania, Philadelphia, Pennsylvania 19104}

\address{$^{3}$ Department of Physics and Astronomy,
         and Center for Fundamental Materials Research, \\
         Michigan State University,
         East Lansing, Michigan 48824-1116}

\date{Received \hspace{3.0cm}}

\maketitle


\begin{abstract}
We show that an electrical   shift current is generated when
 electrons are photoexcited from the valence to 
conduction bands on a BN nanotube.  This photocurrent follows the light pulse envelope 
and its symmetry is controlled by the atomic structure of the nanotube.
 We  find that the shift current has an intrinsic quantum 
 mechanical signature in which the chiral index of the tube  determines
 the direction of the current along the tube axis.
 We identify the discrete lattice effects in
 the tangent plane of the tube that lead to an azimuthal component
 of the shift current. The nanotube shift current  can lead to ultrafast opto-electronic
 and opto-mechanical applications.
\end{abstract}




 ] 


\narrowtext

Recent progress in synthesis of nanometer scale materials has led to 
the discovery of ${\rm B_xC_yN_z}$ nanotubes. They can be formed in
nearly homogeneous ordered structures \cite{Chopra,Weng} or in multiwall 
hybrid  structures with alternating BN and C compositions \cite{NT99BCN}. 
These materials differ fundamentally from the structurally similar 
carbon nanotubes by being {\it non-centrosymmetric} (NCS) and {\it polar}.  This opens the possibility for accessing a 
new class of photovoltaic effects at the molecular scale. 

Photovoltaic effects in NCS materials are often based on an asymmetric
generation of hot carriers at momenta $\pm {\bf k}$ leading to a so 
called ``ballistic photocurrent" \cite{Sturman,KralPUMA,MeleKT}. 
In polar NCS materials, photoexcitation across the bandgap with 
 polarized light also produces 
the so called electrical ``shift current" \cite{Belinicher,Sipe,KralSHI}.
Microscopically this originates from a net displacement of charge
in the unit cell due to light induced interband 
transitions (${\bf J_e}$), intraband relaxation (${\bf J_s}$) and (radiative) 
transitions to the original bands (${\bf J_r}$).  In bulk materials, the 
excitation component ${\bf J_e}$ usually prevails.

We show that {\it unpolarized light} can induce a shift current ${\bf J_e}$ 
in polar NCS nanotubes, with a direction along the tube axis which is
determined by the chiral index of the tube. We find that  
 this effect has an essentially quantum mechanical origin,  where
 the {\it sign} of the current along the tube axis is controlled by the phase matching
 of electronic Bloch waves around the tube circumference. The
  discrete lattice
 structure controls the azimuthal component of this current and thus produces
 a net {\it helical current} on the tube. 
These photo-effects can lead to an assortment of new opto-electronic, 
opto-mechanical and magnetic applications, and  interesting extensions to
ring structures \cite{cropcircles} and heterojunctions \cite{Blase}.
 We should stress that the photoeffects we discuss here are found at  nonlinear order
 in the exciting fields, and are therefore physically distinct from chiral
 currents tilted with respect to 
 the tube axis, predicted in dc bias driven chiral BN nanotubes 
\cite{MiyamotoI} or chiral stretched C nanotubes \cite{MiyamotoII,Yevtushenko}. 

A flat BN sheet has a honeycomb lattice  with the B and N occupying alternate
sublattices as shown in Figure 1. 
  The physics of the shift current can be understood qualitatively by
considering the response of this 2D network of bonds to normally incident polarized light. The valence states
are polarized towards the N sites and the conduction states towards the B sites. For  vertically polarized ($y$-polarized) incident light
the response of the system is dominated by the bonds which produce a net $y$ polarized current. For
  horizontally ($x$-polarized) incident light, the bonds with a 
 nonzero horizontal component dominate, but these  also produce a net $y$-polarized current.
Note that excitation with an unpolarized incident field {\it cannot} produce a net current on  this lattice since it has a threefold
symmetry. However, this symmetry is removed when the sheet is wrapped into a cylindrical nanotube
where 
 the physics should be dominated by excitations with the field polarized along  the tube axis. Depending on the wrapped structure of the tube,  we anticipate   an
 intrinsic photocurrent which can flow along the tube, around the tube, or in a chiral pattern on
 the tube surface.

\begin{figure}[t]
       \epsfxsize=2.5in
      \centerline{\epsfbox[55 354 479 723]{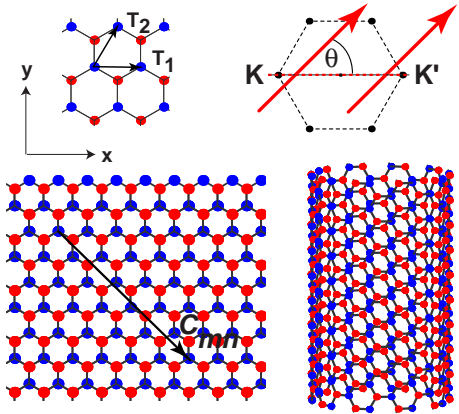} } 
\vspace{5mm}
\caption{Planar BN forms a honeycomb lattice with B and N occupying alternate
sites (upper left). A BN tube is formed by wrapping the sheet through the 
translation vector $C_{mn}$ (lower left) quantizing the transverse 
 crystal momenta.  The upper right hand panel 
shows two representative lines of allowed momenta for a tube with a nonzero
chiral index $\nu = {\rm mod}(m-n,3)$ and chiral angle $\theta$.  
 The lower right panel shows the structure for the nonchiral  armchair (10,10) wrapping of the BN sheet.}
\label{f1}
\end{figure}

These ideas can be quantified by developing a quantum mechanical model which generalizes
 our long wavelength theory for a nonpolar carbon nanotube \cite{kanemele} to
 the heteropolar lattice. We work in a basis of Bloch orbitals $\Phi_{k \alpha}(r)
 = e^{i k \cdot r}
                \sum_n e^{-i k \cdot (r - R_n - \tau_{\alpha})} \phi_{n \alpha}/\sqrt{N}
                = e^{i k \cdot r} U_{k \alpha}$
where $\alpha = \pm 1$ denote the two sublattices occupying sites at $\tau_{\alpha}$ in
the $n$-th unit
 cell.  We study the states near the conduction and valence band edges
 at the K and K' points of the Brillouin zone shown in Figure 1. A long wavelength Hamiltonian for these states
 is obtained by an expansion in small crystal momenta around these points, and yields in
 our two component basis ($\hbar = 1$)
 \begin{equation}
H_{\lambda}(\Delta,k,\delta_{\lambda}) =
\left( \begin{array}{cc} \Delta& \lambda v_F (k - i \delta_{\lambda})
 \\
 \lambda  v_F (k + i \delta_{\lambda}) & -\Delta \end{array} \right)
\label{h2x2}
\end{equation}
 where $\lambda = \pm 1$ is an index which labels an expansion around
 the K or K' points.
The Hamiltonian in Eq. (\ref{h2x2}) is parameterized by three energies:
 a symmetry breaking site diagonal potential $\Delta$, the kinetic energy
 $v_F k$ for motion along the tube, and
 the kinetic energy  $v_F \delta$ due to the quantization of the transverse momentum
 around the circumference of the tube.
  On a cylindrical tube with lattice constant $a$ and
 primitive
lattice translation vectors in its tangent plane $T_1 = a(1,0)$ and $T_2 = a(1/2,\sqrt{3
}/2)$ (Figure \ref{f1}) it is conventional to index the lattice structure
   by two integers $m$ and $n$ which define a superlattice
 translation vector ${\cal C}_{mn} = m T_1 + n T_2$. The  transverse momenta on an $(m,n)$ tube are
 quantized to the values $ \delta_N = \delta_0 + 2 \pi \, N /|{\cal C}_{mn}|$
where  $\delta_0 =   2 \pi \, {\rm sgn}(\nu)/3|{{\cal C}_{mn}}|$, depending
  on the sign of the chiral index of the tube, $\nu = {\rm mod}(n-m,3)$.
 In Eq. (\ref{h2x2}) $\delta_{\lambda} = \lambda \delta_N$.

 The eigenvectors of  $H$ in Eq. (\ref{h2x2}) are
\begin{equation}
g_v =  \left( \begin{array}{c}
                                u e^{-i \phi/2} \\
                               -v e^{ i \phi/2} \end{array} \right)  \frac{e^{i\theta_{vk}}}{\sqrt{2}}
\,\, ; \,\,
g_c =  \left( \begin{array}{c}
                                v e^{-i \phi/2} \\
                                u  e^{ i \phi/2} \end{array} \right)  \frac{e^{i\theta_{ck}}}{\sqrt{2}}
\label{g}
\end{equation}
where $E= \sqrt{v^2_F (k^2 + \delta_{\lambda}^2) + \Delta^2}$, $\phi = {\rm tan^{-1}} (\delta_{\lambda}/\lambda k)$,
  $u=\sqrt{(E-\Delta)/E}$, and $v=\sqrt{(E+\Delta)/E}$. In the rest
 of this paper we will set the Fermi velocity $v_F=1$ so
 that energies and momenta are measured in the same units. Eq. (\ref{g}) explicitly includes  a gauge function
$\theta_{m k}$ since the overall phase of the Bloch function
 is not fixed, and our calculation will require differentiation with respect to $k$. The Bloch eigenfunctions of our problem are
 expressed as the product of three factors
$\psi_{km}(r) = e^{i k \cdot r} \sum_{\alpha} g_{m \alpha}(k) U_{k \alpha}(r)$.

We extend the model Hamiltonian in Eq.(\ref{h2x2}) to include coupling of 
electrons to the oscillating optical fields ${\bf  E}(t)={\bf  E}~
e^{-i \omega t} + c.c.$ through the dipole operator \cite{Sipe,KralSHI}
$H_{int} = - e~{\bf  E}(t) \cdot {\bf  r}$. For a field
 polarized along the tube direction the matrix elements 
of the dipole operator between band eigenstates $m$ and $m'$ at 
crystal momenta $k$ and $k'$ are expressed using a formulation due to 
Blount \cite{Blount}
\begin{eqnarray}
r_{m'm}(k',k)&=& \langle \psi_{k'm'} | r | \psi_{km} \rangle
 =  -i~\langle \psi_{k'm'}| \partial_k |\psi_{km} \rangle \nonumber\\
&+& i~\delta(k-k') \sum_{\alpha\beta} \delta_{\alpha\beta}~
g^*_{m' \beta}(k') \partial_k g_{m \alpha}(k) \ .
\label{rmmkk}
\end{eqnarray}
The second term on the r.h.s of (\ref{rmmkk}) forms the connection 
$\xi_{m'm}= i g^* _{m'}(k) \cdot \partial_k g_m(k)$ due 
to the $k$ dependence of the eigenstates in (\ref{g}) \cite{Gamma}. 

In general the shift current ${\bf J}$ can be expressed in terms
of the  band off-diagonal matrix elements of the
 velocity operator ${\bf v}_{\beta \alpha}$ (${\bf v}=i[H,{\bf r}]$), and the band 
off-diagonal term in the density matrix $\rho_{\alpha \beta}$, 
calculated to second order 
in the exciting fields ${\bf E}$, as follows \cite{Belinicher,Sipe,KralSHI}
\begin{equation}
{\bf J}  = 2 {\rm e} \sum_{m\neq n } \int \frac{ dk }{2 \pi}~ 
{\bf v}_{mn}(k)~\rho_{nm} (k)={\bf J}_e+{\bf J}_s+{\bf J}_r\ .
\label{trace}
\end{equation}
We focus on the component of the excitation 
current $ {\rm J_e}$ {\it along the tube direction}. After evaluating the sum in (\ref{trace}), we arrive at 
the transparent result 
\begin{equation}
{\rm  J_e }= 2e \int \frac{ dk }{2 \pi}~ \dot f_{cv}(k)~R_{cv}(k) 
\approx e~\dot{n}~{\bf \cal R}\ .
\label{sc}
\end{equation}
Here $\dot f_{cv}(k)$ is the transition rate at wave vector $k$,
and the {\it shift vector} $R_{cv}(k)$ is given by
\begin{equation}
R_{cv}(k) = \partial_{k} \theta_{vc}(k) + \xi_{cc}(k) - \xi_{vv} (k)\ 
\label{sv}
\end{equation}
with $\xi_{vc}(k) =|\xi_{vc}(k)|\ e^{i \theta_{vc}(k)}$.
The shift vector $R_{cv}(k)$ is invariant under the gauge transformations
${\rm exp}(i \theta_{mk})$ and we will evaluate it in a gauge with
$\partial_k \theta_{mk} = 0$.

For an incident electric field polarized along the tube direction, interband 
excitations are allowed only between band states with the same transverse 
momentum $\delta$. We consider transitions between the lowest 
two azimuthal subbands  ($N=0$).  Then using the eigenfunctions 
in Eq.(\ref{g}) we find  
\begin{equation}
\xi_{vc}(k) = \frac{-1}{2} uv \left( \frac{ \Delta}{E^2 - \Delta^2} 
\partial_k E + i \partial_k \phi \right)\ ,
\label{rcv}
\end{equation}
which gives the off-diagonal contribution in Eq. (\ref{sv})
\begin{equation}
\partial_k \theta_{vc}  =  \partial_k {\rm tan^{-1}} \left( \frac{- \delta 
\sqrt{\delta^2 + \Delta^2 + k^2}}{k \Delta} \right)\ .
\label{svc}
\end{equation}
The diagonal elements in  Eq. (\ref{sv}) are 
\begin{eqnarray}
\xi_{mm}(k) = i \,\,g^{\dagger}_{m} \cdot \partial_k g_{m} 
  = \mp \frac{\Delta}{2E} \left( \frac{\delta}{\delta^2 + k^2} \right)\ ,
\label{rccvv}
\end{eqnarray}
where the sign is negative for valence states and positive for conduction
states.  Combining Eq. (\ref{svc}) and Eq. (\ref{rccvv}) we obtain 
the shift vector
\begin{equation}
R_{cv}(k) =  \frac{ 2 \delta \Delta}{(\delta^2 + k^2)
\sqrt{\delta^2 + \Delta^2 + k^2}}\ .
\label{shift}
\end{equation}
Comparison with Eq. (\ref{rccvv}) shows that the shift current  (Eq.(\ref{sc})) opposes  the direction 
of the ground state polarization of the tube.

\begin{figure}[t]
          \epsfxsize=2.0in
         \centerline{\epsfbox[33 351 571 730]{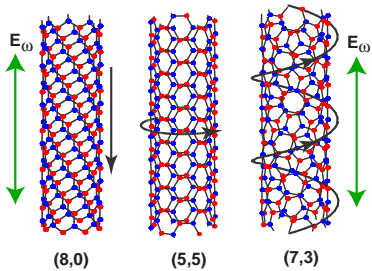} } 
\vspace{5mm}
\caption{Electronic shift currents produced by excitation of BN nanotubes
with light polarized along their axis. For the (8,0) wrapping the
chiral index is $\nu = -1$, and the shift current is purely longitudinal. 
 The (5,5) armchair wrapping has chiral index $\nu=0$ and has zero longitudinal current and a nonzero azimuthal
 current which runs ``counterclockwise" on the tube surface. The (7,3) wrapping
 has a chiral shift current with nonzero longitudinal and azimuthal components.}
\label{f2}
\end{figure}

This shift vector is odd in the symmetry breaking potential $\Delta$ 
and {\it odd in the transverse momentum} $\delta$. Therefore, two BN tubes with nearly 
the same wrapped lattice structure but {\it opposite} chiral indices 
$\nu$ exhibit reversed ground state polarizations and opposite 
shift currents. Armchair BN nanotubes with wrapping indices $(m,m)$
have chiral index $\nu = 0$ and do not exhibit a longitudinal shift 
current (as shown in the middle panel in Figure \ref{f2} for the (5,5) tube).  
Zigzag BN tubes with wrapping indices $(m,0)$ can be grouped into three 
families with positive, negative or zero shift currents distinguished 
by the sign of the chiral index $\nu$ (an example for the (8,0) tube 
with $\nu = -1$ is in the left panel of Figure \ref{f2}). 
 It is important to notice that
 an isolated flat BN sheet has a threefold symmetry axis perpendicular
 to the BN plane and  therefore it has zero electric polarization in
 its tangent plane, by symmetry. Thus it 
 has no shift current for excitations at the band edge. The nonzero shift vector
 in Eq. (\ref{shift}) when $k \rightarrow 0$
  is a remarkable long wavelength quantum mechanical effect that reveals  the 
 quantization of the transverse momentum on the wrapped tube.

The shift current ${\rm J_e}$ in Eq. (\ref{sc}) is determined by the average shift vector
${\cal R}$, describing the change of polarity in the excitation, and by the carrier 
injection rate $\dot{n}$. The BN bond has significant ionic character with 
\cite{Rubio} 2$\Delta \approx$ 4 eV.  This gives for excitations near the 
band edge, $k \rightarrow 0$, a shift vector with a magnitude 
$ {\cal R} \approx 0.06 \, 
{\rm nm}$ for a (17,0) tube.  For an incident intensity ${\rm S 
= 100 \,\, kW/cm^2}$ we can also obtain the interband excitation rate; 
we estimate that $\dot n \approx{\rm  70 \,\,  nm^{-1} \mu s^{-1}}$ produces a net electronic shift
 current of  ${\rm J_e} \approx$ 0.67 pA. This value is slighly larger than one
 obtains for excited bulk polar semiconductors.  In a coaxial multiwall tube $\rm{R_{cv}}$
  can change  magnitude and sign on each wall. Although opposing currents
  on different tube walls reduce the macroscopic current
  a complete cancellation is very unlikely since the  magnitude of $\rm{R_{cv}}$ depends on the tube
  radius. We also expect that the recombination current ${\rm J_r}$ will oppose
${\rm J_e}$ since both currents rely on optical transitions with longitudinal polarization. However, this should
 not prevent the observation of ${\rm J_e}$ using pulsed laser excitation since the timescale for
 ${\rm J_r}$ is of order nanoseconds. 

This formalism can be extended to study {\it chiral shift currents} 
with nonvanishing longitudinal and azimuthal components. 
Interestingly, the long wavelength Hamiltonian in Eq. (\ref{h2x2}) gives
zero azimuthal current for {\it any} chiral index $\nu$.  As $Q_{\pm} 
=  k \pm  i\delta \rightarrow 0$  this occurs because of a cancellation 
of contributions from the K and K' points whose long wavelength 
Hamiltonians have opposite handedness. However, azimuthal currents 
arise due to discrete lattice effects occurring at higher orders in an 
expansion in $Q_{\pm}$.  To study this, we define two unit vectors at each point in the tube 
surface, perpendicular to the local outward unit normal $\hat n$: 
$\hat e_x$ which points along the tube axis, and $\hat e_y = \hat n 
\times \hat e_x$ which points ``counterclockwise" on the tube surface.  
A chiral angle $\theta $ is defined as the angle between $\hat e_x$ and the $T_1$ lattice 
direction as shown in Figure \ref{f1} (with this
 convention armchair nanotubes have $\theta$ = 0). 
Then the lowest nonvanishing contribution to an azimuthal ($y$-polarized) 
shift current 
must have the form 
$\cos (6 \phi - 3 \theta)$ where $\phi = \tan^{-1}(\delta/k)$. 
 The dependence on 
$6 \phi$ means that azimuthal currents first occur in the continuum 
theory at sixth order in $|Q|$. 

We can study these terms by replacing the kinetic energy terms  in the Hamiltonian of Eq. 
(\ref{h2x2}) by the discrete lattice counterpart 
$iv_F (k + i \delta) =i v_F Q_+ \rightarrow t \sum_n e^{i (K + Q) \cdot \tau_n}\ 
$
where $t$ is the hopping amplitude between sites connected by 
the bond vectors $\tau_n$.  A systematic expansion  of the azimuthal
 shift current in  powers of $Q_{\pm}$ 
then gives contributions with the correct lattice symmetry and shows 
that the ratio of the azimuthal and longitudinal in-plane current densities
scales as
\begin{equation}
 p = \frac{J_y}{J_x} = \frac{ (k^2 + \delta^2)^3 \tau^3}
{\delta (\Delta^2 + k^2 + \delta^2)} \cos(6 \phi - 3 \theta) \, .
\label{pitch}
\end{equation}
The azimuthal current for any zigzag wrapping is zero because of the 
angular factor,  while its longitudinal current, shown in Figure \ref{f2}, 
is determined by the chiral index $\nu$. For armchair tubes and 
all non-zigzag tubes with chiral index $\nu=0$, we have $\delta=0$,
so the shift current is purely azimuthal with a direction determined 
by the sign of the potential $\Delta$. These energy dependent 
currents vanish at the band edge as $\propto (E^2 - \Delta^2)^3$. 
Non-zigzag tubes with chiral index $\nu \ne 0$ have a  nonzero  
$\delta$ and thus the azimuthal current is nonzero with $ p \propto 
(\delta/\Delta)^2 (\delta \tau)^3$. There the shift current circulates in 
a {\it helical pattern} on the tube surface as shown in the right panel 
of Figure \ref{f2}, with a chirality determined by the index $\nu$.  
The pitch is also energy dependent and increases according to $ p(E) =  p(0) 
+ a (E^2 - \Delta^2)$. Thus, as shown in Figure \ref{f2}, the zigzag nanotube
generate a photocurrent like a wire, the armchair tubes like a coil, 
and the chiral tubes can exhibit a helical current on the 
tube surface.

 Alternatively, the shift current can be studied in a semiclassical model, by
 considering an expansion in $t/\Delta$, where $t$ is the nearest neighbor
 hopping amplitude, and $\Delta$ is the site diagonal potential.
 In the limit $t/\Delta << 1$  the tube can be regarded as a network of independent bonds whose excitations are superimposed.
 Here the local symmetry of the zigzag tube requires zero net azimuthal current while the armchair
 wrapping has a nonzero azimuthal current with its direction  determined by the
  the sign of   $\Delta$.  However, it is difficult to correctly 
describe  the physics of the {\it longitudinal} current
 using this model. This is because the longitudinal current is controlled
 by the quantization of the  transverse momentum $\delta$. The first
 terms in the semiclassical expansion that are sensitive to $\delta$ occur
 at order $(t/\Delta)^N$ where $N$ is the number of bonds ($\approx 40$) around
 the tube circumference. This reflects the fact that the longitudinal current
 is fundamentally a nonclassical quantity for this system.

The longitudinal shift current  
 reduces the ground state electric polarization of the nanotube.
   This can lead to a 
{\it fast readjustment} of atomic positions on the tube walls. 
 Generation of voltages by a mechanical elongation 
\cite{Salvetat} of NCS nanotubes is a related effect. 
 In both situations the response can be stronger in buckled nanotubes \cite{Hernandez} or if 
different types of atoms occur in  different layers, as in MoS$_2$, 
WS$_2$ \cite{Tenne} or GaSe nanotubes \cite{Cote}. 
Recently, mechanical motions, strikingly similar to 
those found in some biomolecular systems
\cite{Samuel}, were observed in irradiated nanotube bundles \cite{Zhang}.
Interesting mechanical response to an applied dc bias,
is observed in carbon nanotubes immersed in a solution of NaCl \cite{Baughman}.
Here, the motion is caused by {\it attraction of ions} of a given polarity 
to the nanotube. The speed but short duration of the shift current-induced mechanical response 
 complements the slow but steady state effect of a dc bias. 

Experimental observation of the shift current in BN nanotubes  would provide a striking illustration of 
fundamental concepts in the 
modern quantum theory of polarization. The effect is  also promising for new applications since it couples  the microscopic physics
 on short length scales, which can be tuned by local fields
 and mechanical loads, to the long distance  properties of the system (polarization, photovoltage, etc.)  Finally, there is a rapidly growing family of
 related submicron one dimensional materials  to which these
 ideas can be applied. 

\vspace{3mm}
\noindent
{\bf Acknowledgments}\\
PK acknowledges J.E.~Sipe and M. Shapiro for financial support. This work was 
also supported by the DOE under Grant number DE-FG02-84ER45118 (EJM)
 by the NSF under Grant number DMR 98-02560 (EJM) and by the ONR under Grant Number N00014-99-1-0252 (DT).




\begin{references}
\bibitem{Gamma} 
Eq.(\ref{rmmkk}) is gauge invariant if $ i\delta_{\beta \alpha} 
\partial_k~g_{m \alpha}(k)$ is generalized to the covariant derivative 
$(i\delta_{\beta \alpha} \partial_k + \Gamma_{\beta \alpha}(k))~ 
g_{m \alpha}(k)$, where $\Gamma_{\beta \alpha}(k) = i \int 
dr~U^*_{k \beta}(r) \partial_k U_{k \alpha}(r) $ arises from the 
$k$-dependence in the basis functions $U_{k \alpha}(r)$. Our choice 
of basis in Bloch functions gives $\Gamma_{\alpha \beta}(k) = 0$.
\end{references}
 \end{document}